\documentclass[twocolumn,prl,showpacs]{revtex4-1}
\usepackage{graphicx,color}
\usepackage{dcolumn}
\usepackage{amsmath,amssymb}
\usepackage{bm}

\newcommand\Cu{0.55}

\newcommand{\MeijerG}[2]{%
  G_{1,5}^{4,1}
  \Bigl(#1\Bigl|\genfrac{}{}{0pt}{}{\lower0.8mm\hbox{0}}
       {\raise0.3mm\hbox{#2}}\Bigr)
}

\newcommand\figurescalesmall{0.39}
\newcommand\figurescalelarge{0.87}

\newcommand\ZUichi{%
  \begin{figure}[t]
   \includegraphics[width=\figurescalesmall\linewidth]{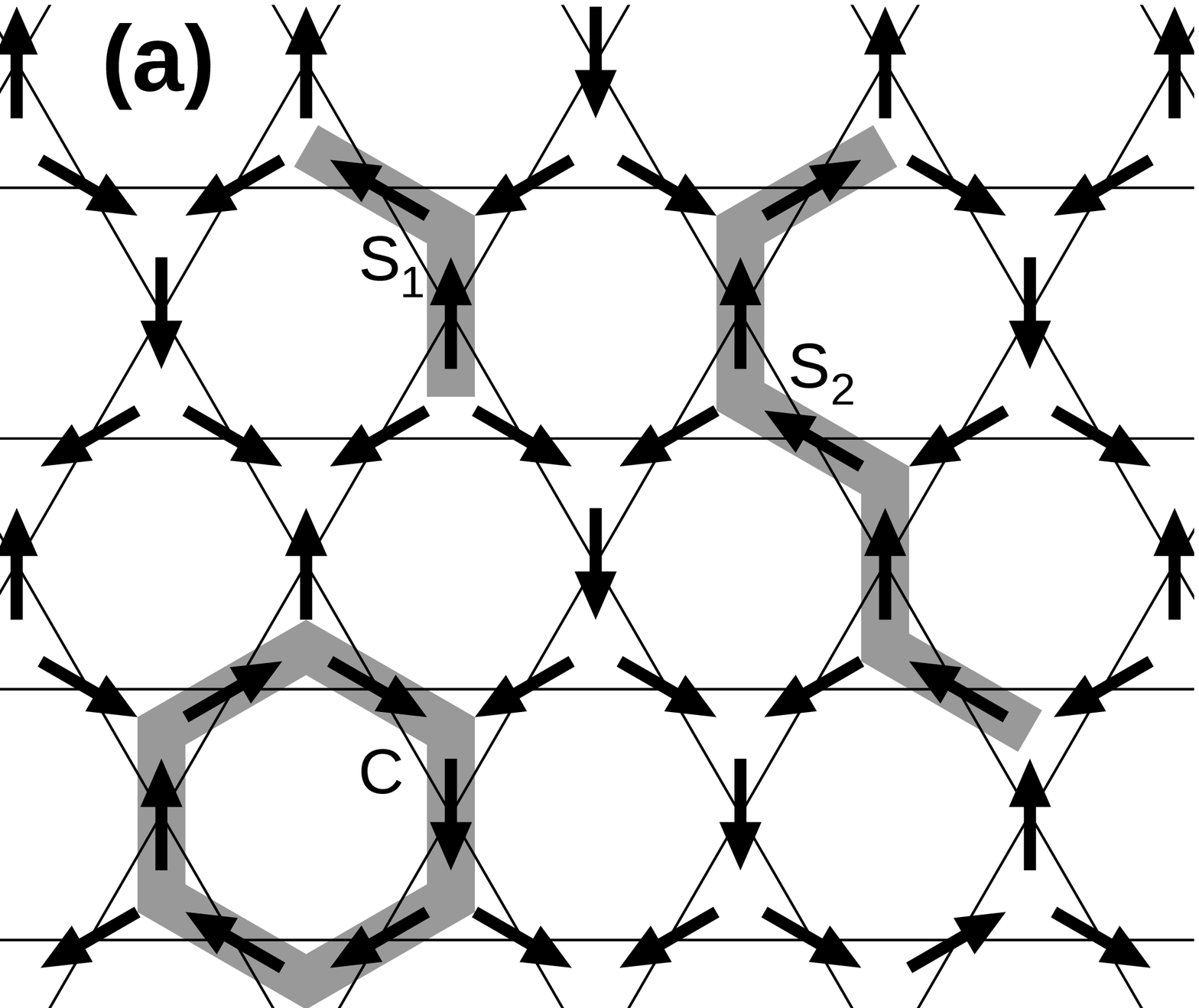}
   \hspace{4.3mm}
   \includegraphics[width=\figurescalesmall\linewidth]{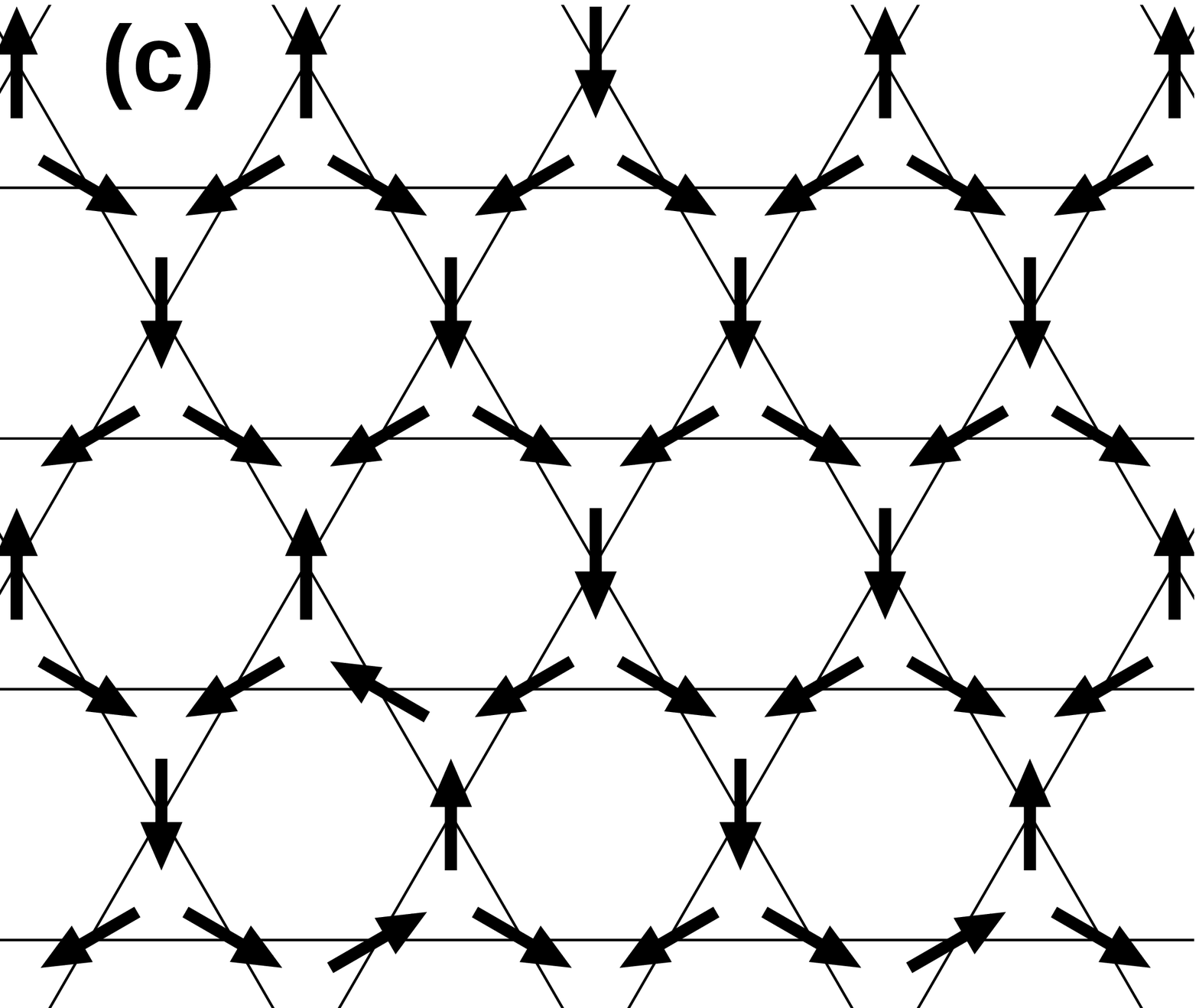}

   \vspace{3.0mm}
   \includegraphics[width=\figurescalesmall\linewidth]{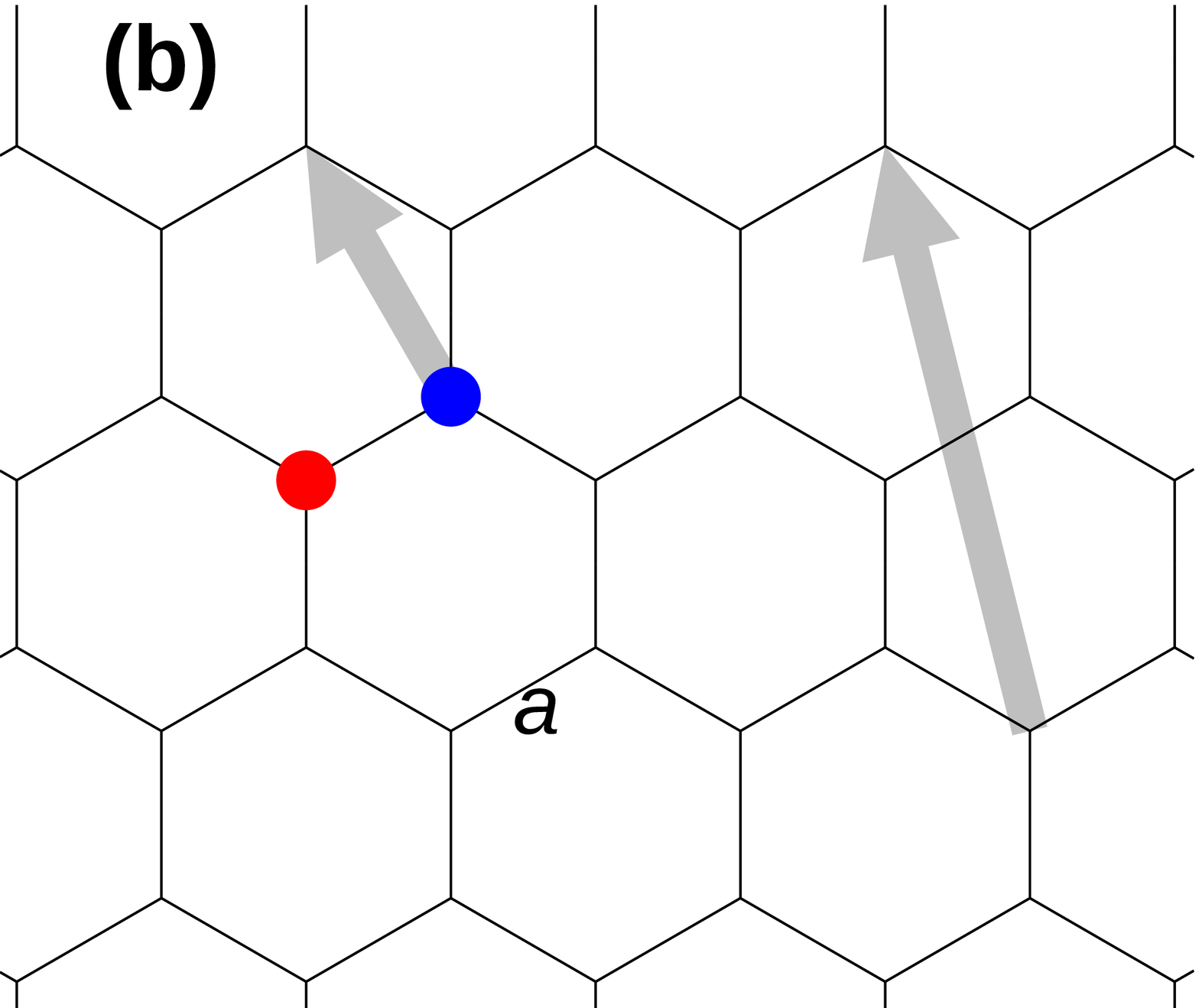}
   \hspace{4.3mm}
   \includegraphics[width=\figurescalesmall\linewidth]{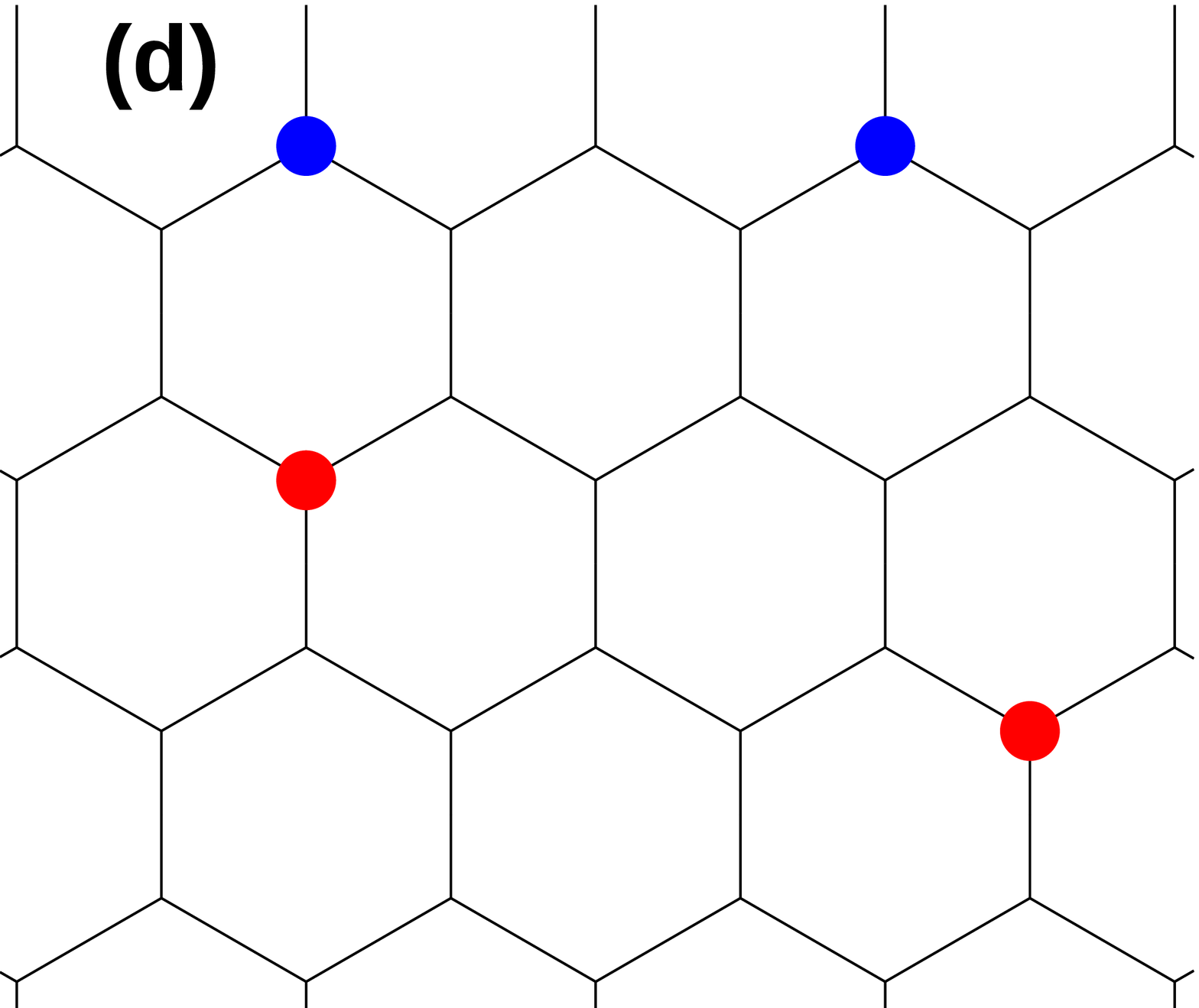}
   \caption{%
   (Color online)
   Spin (monopole) configurations on $\Lambda^*$ ($\Lambda$)
   are given in (a) and (c) [(b) and (d)].
   (a)
   The loop (${\cal C}$) and strings (${\cal S}_{1,2}$) are given
   by thick gray lines along which spins line up tail to nose.
   (b)
   The red (blue) circle represents a positive (negative) charge
   monopole; $a$ indicates the site spacing.
   Thick gray arrows show the displacements.}
   \label{MP}
  \end{figure}}

\newcommand\ZUni{%
  \begin{figure}[t]
   \includegraphics[width=\figurescalelarge\linewidth]{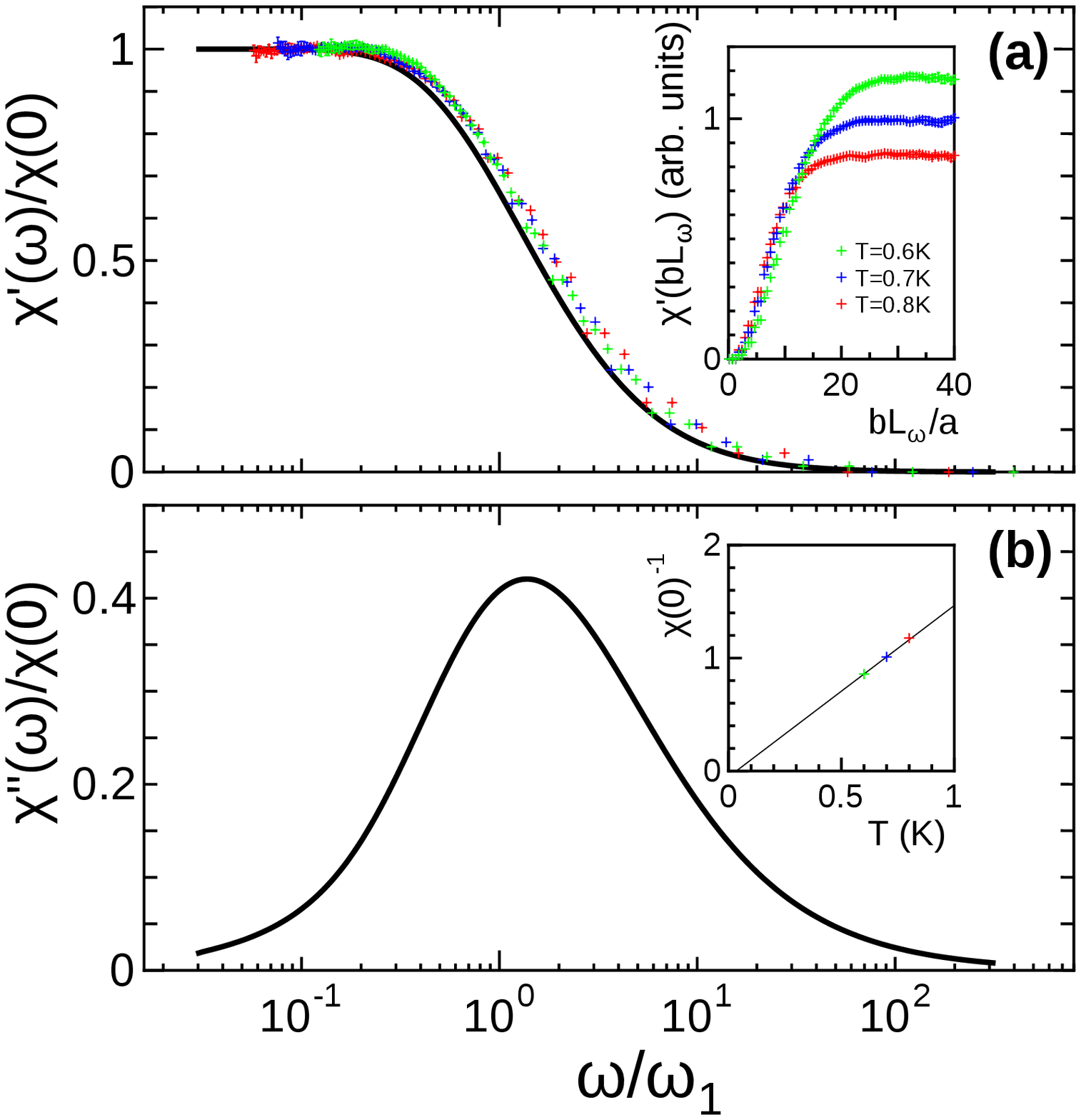}
   \vspace*{-1mm} 
   \caption{%
   (Color online)
   The ac susceptibility:
   (a) Eq.\ (\ref{ReX_result}) and  (b) Eq.\ (\ref{ImX}).
   The frequency is given in units of $\omega_1$, and is plotted on a
   logarithmic scale, where $b=\frac14$ (see text).
   The cut-off dependence of $\chi'$ obtained by MC simulations are
   given in the inset of (a); the value $1/\chi(0)$ is plotted in the
   inset of (b), where a fitting line is also given by using data at the
   lowest two temperatures.
   The scaled MC simulation data using monopole density and $\chi(0)$
   are given in (a).}
   \label{acX}
  \end{figure}
}

\newcommand{\Dy}{Dy$_2$Ti$_2$O$_7$}
\newcommand{\Ho}{Ho$_2$Ti$_2$O$_7$}
\newcommand{\kBT}{k_{\rm B}T}
\newcommand{\nm}{n_{\rm m}}
\newcommand{\etal}{{\it et al.}}

\def\TITLE{%
A Scaling Theory for ac Magnetic Response in Kagom\'e Ice}
\def\AUTHORS{%
Hiromi Otsuka, Hiroshi Takatsu, Kazuki Goto, and Hiroaki Kadowaki}
\def\AFFILATION{%
Department of Physics, Tokyo Metropolitan University, Tokyo 192-0397, Japan}

\begin{document}

\title{\TITLE}
\author{\AUTHORS}
\affiliation{\AFFILATION}
\date{\today}
   
\begin{abstract}
 A theory for frequency-dependent magnetic susceptibility $\chi(\omega)$
 is developed for thermally activated magnetic monopoles in kagom\'e
 ice.
 By mapping this system to a two-dimensional (2D) Coulomb gas and then
 to a sine-Gordon model, we have shown that the susceptibility has a
 scaling form
 $\chi(\omega)/\chi(0)={\cal F}(\omega/\omega_1)$,
 where the characteristic $\omega_1$ is related to a charge correlation
 length between diffusively moving monopoles, and to the sine-Gordon
 principal breather.
 The dynamical scaling is universal among superfluid and superconducting
 films, and 2D XY magnets above Kosterlitz-Thouless transitions.
\end{abstract}


 \maketitle

 Frustrated spin systems have attracted considerable attention for
 decades, because they provide an opportunity of uncovering novel phases
 and excitations
 \cite{Lacrox11}. 
 Even in the simplest cases of the Ising antiferromagnets on triangular
 \cite{Wann50}  
 and kagom\'e
 \cite{Shozi51}
 lattices, exact solutions revealed the absence of magnetic order and
 the macroscopic degeneracy in the ground states, which are viewed as
 the hallmark of the frustration.

 Rare-earth pyrochlore oxides such as {\Ho}
 \cite{Harr97}
 and {\Dy}
 \cite{Rami99}
 proffers a new paradigm in this research area
 \cite{Bram01}.
 It is considered that
 despite large magnetic moments $\mu_{\rm eff}\sim 10\mu_{\rm B}$
 the spins on the pyrochlore lattice do not order down to a quite low
 temperature $\sim 0.1$ K
 \cite{Harris98b,Bram01},
 and exhibit a residual entropy
 \cite{Rami99}
 (although possibilities of a magnetic order
 \cite{Melk01}
 and an absence of the residual entropy
 \cite{Poma13}
 have been reported).
 The origin of these behaviors can be attributed to a strong Ising
 anisotropy with respect to the local $\langle111\rangle$ axis and an
 effective ferromagnetic coupling between neighboring spins
 \cite{Hert00},
 which, then, force the spins at four corners of each tetrahedron to
 satisfy the two-in and two-out condition.
 Since this constraint is the same as that for proton configurations in
 water ice
 \cite{Bern33},
 those materials are named as ``spin ice''.

 Recently,
 point-defect excitations in spin ice created by breaking the ice rule
 \cite{Ryzh05} 
 have been intensively investigated
 \cite{Kado09,Morris09,Fennell09,Bovo13}, 
 since the intriguing prediction of magnetic monopoles
 \cite{Cast08,Cast11}.
 These excitations behave as quasi-particles with magnetic charges
 moving on the diamond lattice 
 \cite{Ryzh05,Bram09,Gibl11}
 like ion defects, H$_3$O$^+$ and HO$^-$, in water ice
 \cite{Bjer51}.
 While much efforts have been paid to account for their static and
 dynamical properties, there still exist unclear points and subjects to
 explore
 \cite{Bram12}.
 This is partly because the monopole-like excitation is a topological
 defect, and is a nonlocal object emerging in the vicinity of the
 ground-state manifold.

 A way to circumvent its intractability is to make it move in more
 restricted space, e.g., in two dimensional (2D) space.
 A 2D spin ice can be achieved by applying a magnetic field
 ${\bm H}_{\rm dc}$ along a [111] direction, along which the pyrochlore
 lattice is stacking of triangular and kagom\'e lattices
 \cite{Harr98,Mats02}. 
 When the [111] field is not very high, the spins on the triangular
 layers are fixed parallel to the field direction at low temperatures,
 and consequently the spins on each kagom\'e layer are decoupled
 and remain frustrated, which endowed the name ``kagom\'e ice''
 \cite{%
 Mats02,Saka03,Hiroi03,Higashinaka04,Taba06,Fennell07,Uda02,Moes03,Isak04}.
 The kagom\'e-ice state is characterized by a magnetization plateau
 \cite{Mats02,Saka03}
 and a reduced residual entropy
 \cite{Mats02,Hiroi03,Higashinaka04,Uda02,Moes03}.
 In the low temperature limit $T \rightarrow 0$, it is in a Coulomb
 phase
 \cite{Henley09} 
 with the power-law decay of spin correlations
 \cite{Moes03}. 
 At low $T$,
 kagom\'e ice is characterized by a long charge correlation length
 $\xi$ or a small monopole density $\nm\propto\xi^{-2}$
 \cite{Moes03,Taba06,Fennell07}.

 A minimal Hamiltonian for kagom\'e ice is a nearest-neighbor (NN)
 model
 \cite{Harr98} 
 consisting of one kagom\'e layer and neighboring two triangular layers
 with pinned spins:
 \begin{equation}
  H_{\rm ice}
   =J_{\rm eff}
   \sum_{\langle i j \rangle} \sigma_i \sigma_j 
   -\mu_{\rm eff}
   \sum_i {\bm H}_{\rm dc}\cdot\hat{\bm z}_{a(i)}\sigma_i~~
   (J_{\rm eff}>0).
 \end{equation}
 $\sigma_i$ ($=\pm1$) is an Ising variable for a spin
 ${\bm S}_i=\sigma_i\hat{\bm z}_{a(i)}$
 at a site $i$ on a sublattice $a(i)$, and
 $\hat{\bm z}_{a(i)}$
 stands for a unit vector parallel to the local Ising axis.
 The NN exchange interaction is antiferromagnetic in terms of Ising
 variables.
 The ice rule, requiring $\sum_{a=1}^4\sigma_a=0$ for each tetrahedron,
 can be broken by thermal activation, creating magnetic monopoles with a
 charge $q=\frac12\sum_{a=1}^4 \sigma_a=\pm1$, which is illustrated in
 Fig.\ \ref{MP}.

 Another motivation of this work originates from our recent experimental
 studies on the dynamics of monopoles moving in the kagom\'e plane of
 {\Dy}
 \cite{Takatsu13}: 
 As pointed out there, 
 the 2D dynamics of monopoles excited from the kagom\'e-ice state can be
 investigated by applying an ac magnetic field 
 ${\bm H}_{\rm ac}(t)={\rm Re}[{\bm H}_0 e^{i\omega t}]$ 
 perpendicular to
 ${\bm H}_{\rm dc}$, which works as driving force for
 monopoles and measuring induced magnetization
 ${\bm M}_{\rm ac}(t)={\rm Re}[\chi(\omega){\bm H}_0 e^{i\omega t}]$.
 The frequency-dependent ac susceptibility $\chi(\omega)$ has been
 indeed measured.
 In the low-frequency ranges of the experiments
 \cite{Snyder01,Matsu01,Takatsu13}, 
 the monopole motion is thought to be governed by diffusion
 \cite{Jaub09},
 characterized by a diffusion constant ${\cal D}$.

 In this letter,
 we have theoretically studied $\chi(\omega)$, and have found that
 it gives deep insight into the monopole dynamics in the kagom\'e-ice
 state.
 Using the monopole degrees of freedom, kagom\'e ice can be mapped to
 the 2D Coulomb gas model
 \cite{Minn87}, 
 in which monopoles interact via the logarithmic Coulomb potential,
 which is caused by the entropic interaction in kagom\'e ice
 \cite{Moes03,Isak04}. 
 The dynamical properties of the 2D Coulomb gas were closely discussed
 \cite{Ambe78,Minn87},
 as the vortex dynamics in the superfluid films
 \cite{Bish77}.
 Based on the sine-Gordon theory
 \cite{Minn87,Poly77},
 we have shown that $\chi(\omega)$ is expressed by a scaling form 
 \begin{equation}
  \chi(\omega)/\chi(0)\simeq{\cal F}(\omega/\omega_1),
 \end{equation}
 where $\omega_1 = {\cal D}/\xi^2$ is a characteristic frequency of
 kagom\'e ice (see below).
 It is expected that
 the scaling function ${\cal F}$ shown in Fig.\ \ref{acX} can be
 used to analyze experimental data on the 2D spin ice systems.
 We note that this scaling is universal in the sense that it is
 applicable not only for kagom\'e ice, but also for other systems:
 superfluid and superconducting films
 \cite{Bish77},
 2D XY magnets above the Kosterlitz-Thouless transition
 \cite{Kosterlitz,Kost74},
 and generic 2D ices possibly including the artificial spin ice
 \cite{Wang06,Mengotti11,Farhan13,Wills02,Chern11,Moller11}.

 \ZUichi

 First,
 we address Ryzhkin's argument which provides a link between the
 magnetization
 ${\bm M}=\mu_{\rm eff}\sum_{l^*\in\Lambda^*}{\bm S}_{l^*}$
 and a polarization of the monopole charge distribution
 ${\bm P}=(2\mu_{\rm eff}/a)\sum_{l\in\Lambda} q_{l}{\bm x}_{l}$
 \cite{Ryzh05}.
 The sum runs over all sites $l^*$ ($l$) in the kagom\'e (honeycomb)
 lattice $\Lambda^*$ ($\Lambda$).
 The factor, $2\mu_{\rm eff}/a$, is for later convenience, and $a$ is
 the site spacing of $\Lambda$.
 Note that, we henceforth focus on one kagom\'e layer.
 In Figs.\ \ref{MP}(a) and \ref{MP}(b),
 we sketch a spin configuration and the corresponding monopole charge
 distribution.
 The monopole on the ${\cal A}$ (${\cal B}$) sublattice is viewed as a
 particle with positive (negative) charge $q_l=\pm1$.
 To demonstrate a link, we consider changes in ${\bm M}$ and ${\bm P}$
 caused by a directed loop flip of spins along ${\cal C}$ and string
 flips of spins along ${\cal S}_1$ and ${\cal S}_2$.
 While the loop flip (and also any loop flips) does not change them,
 the string flip along ${\cal S}_1$ makes the monopole hop from one end
 to another [see Figs.\ \ref{MP}(b) and \ref{MP}(d)].
 Since the displacement is given by the vector sum of spins associated,
 the change in ${\bm M}$ equals to that in ${\bm P}$, due to the front
 factor.
 Moreover,
 the same relation, $\delta{\bm M}=\delta{\bm P}$, holds for the string
 flip along ${\cal S}_2$ which newly creates a pair of monopoles
 connected by ${\cal S}_2$ [see Fig.\ \ref{MP}(d)].
 The time derivative of the polarization equals to the monopole current
 ${\bm J}=\dot{\bm P}$,
 so these observations lead to Ryzhkin's relation
 ${\bm J}=\dot{\bm M}$.
 In order further to explore their link, let us consider the vacuum of
 monopoles.
 Although ${\bm P}=0$ by definition, there exist spin configurations
 with ${\bm M}\ne0$ belonging to nonzero winding-number sectors.
 In spite of this, we shall approximate ${\bm M}\simeq{\bm P}$ because
 the volume of spin configurations with ${\bm M}=0$ which includes the
 maximally-flippable state
 \cite{Moes00}
 is expected to comprise a major portion of the ground-state manifold.
 Therefore,
 we will focus on a magnetic response accompanied by creations,
 annihilations and rearrangements of monopoles:
 $\chi=\frac{1}{2{\kBT}}\langle {\bm M}^2\rangle_{\rm ice}
 \simeq\frac{1}{2{\kBT}}\langle {\bm P}^2\rangle_{\rm ice}$
 \cite{Jaub12}.

 Since kagom\'e ice exhibits an isotropic charge correlation, we rewrite
 $\chi$, by introducing the charge-density distribution function
 $\rho({\bm x})=\sum_{l\in\Lambda}q_{l}\delta({\bm x}-{\bm x}_{l})$, 
 as
 \begin{equation}
  \chi
   =
   \Omega\frac{\pi\mu^2_{\rm eff}}{{\kBT}a^2}\int_0^\infty dr~r^3 C(r),
   \label{X}
 \end{equation}
 where $r=|{\bm x}|$ and $\Omega$ is the area of $\Lambda$.
 The charge correlation function was defined as
 $C(r)=-\langle\rho({\bm x})\rho({\bf 0})\rangle_{\rm ice}$.

 Equation\ (\ref{X}) exhibits the magnetic response in terms of monopole
 degrees of freedom.
 However,
 it is restricted to the static case, so its extension to the dynamical
 case is our next task.
 For this purpose,
 second, we address Ambegaokar's argument, which provides a link between
 the ac response and a static charge correlation
 \cite{Ambe78}.
 In the analysis of the superfluid film on the oscillating substrate,
 Ambegaokar {\etal} supposed a diffusive motion of vortices, and then
 obtained the ac response by focusing on a role of the mean diffusion
 length during a period of an ac field
 $L_\omega=\sqrt{{\cal D}/\omega}$.
 One intuitive reasoning is as follows:
 Consider a pair of monopoles with separation $r$, and suppose its
 time-dependent polarization keeping {\it in phase} with the ac field.
 Then,
 a monopole should move a distance of order of the separation during one
 ac period, while the mean diffusion length gives a reachable distance
 by diffusive motion within one period.
 Thus, 
 in order to give the in-phase response, the pair should satisfy a
 condition $r<O(L_\omega)$
 \cite{Minn87}.
 Here,
 we also assume a diffusive motion and apply this heuristic argument to
 express the in-phase component (real part) $\chi'(\omega)$ of
 $\chi(\omega)$.
 The result is given by replacing the upper bound of the integral in
 Eq.\ (\ref{X}) with the length of $O(L_\omega)$:
 \begin{equation}
  \chi'(\omega)
   \simeq
   \Omega\frac{\pi\mu^2_{\rm eff}}{\kBT a^2}\int_0^{bL_\omega} dr~r^3 C(r).
   \label{ReX}
 \end{equation}
 We have introduced a constant $b$ less than order of unity.
 In order to infer the value of $b$, we assume a uni-dimensional motion
 of monopoles parallel to ${\bm H}_{\rm ac}$.
 Then,
 the round-trip path length of monopoles is about $4r$ so that the
 condition $4r<L_\omega$ is satisfied.
 We thus assume $b=\frac14$
 \cite{AmbeTeit79}.
 The imaginary part $\chi''(\omega)$ is obtained by applying the
 Kramers-Kronig (KK) relation to $\chi'(\omega)$.
 The above formula is doubly important: 
 It gives the magnetic response in terms of the charge correlation,
 which is possible only for magnets to afford their defect
 representations like the monopole system for kagom\'e ice.
 Also, 
 since the frequency dependence is introduced as
 {\it a finite-size effect}, it may be governed by a ratio of $L_\omega$
 to the characteristic length $\xi$ in kagom\'e ice.
 Below,
 one can see that this is indeed an origin of a scaling nature of
 $\chi(\omega)$.

 Now,
 we can obtain $\chi(\omega)=\chi'(\omega)-i\chi''(\omega)$ via $C(r)$,
 which is given as an average with respect to $H_{\rm ice}$.
 This is rather convenient for numerical calculations; in fact, we
 perform Monte Carlo simulations to evaluate Eq.\ (\ref{ReX}).
 However, to analytically evaluate $\chi(\omega)$, a monopole
 representation of $H_{\rm ice}$ is necessary.
 It has been argued that a gaseous model gives its effective description
 \begin{equation}
  H_{\rm CG}
   =
   \kBT\sum_{l,m\in\Lambda; l>m}-q_lq_mg(|{\bm x}_l-{\bm x}_m|)\kappa,
   \label{CG}
 \end{equation}
 where a neutrality condition $\sum_{l\in\Lambda}q_{l}=0$ is imposed
 \cite{Isak04}.
 $g(r)$ is a lattice propagator to give the correlation between two
 monopoles with a separation $r$ in the ground-state manifold, and
 represents an entropic interaction.
 Since we are focusing on the system with a long correlation length
 $\xi\propto1/\sqrt\nm$, the interaction can be approximated by its
 asymptotic behavior $\ln(r/a_0)$, where $a_0$ is a monopole core
 radius. 
 In this dilute-gas regime, we can obtain a simple universal system by
 coarse-graining a lattice structure and neglecting short-range
 fluctuations
 \cite{Kost74}: 
 Equation\ (\ref{CG}) defines the 2D coulomb-gas (CG) model
 \cite{Kadanoff},
 where $\kappa$ is the inverse CG temperature, and is independent of $T$
 ($\kappa\simeq\frac12$ for an ice-rule system \cite{Fish63}).

 A large reduction of the problem has been attained, but as its
 drawbacks, we should explicitly control a number of monopoles.
 Let us write a $N$-monopole partition function as $Z_N$,
 then the grand-partition function is given as
 $\Xi_{\rm CG}=\sum_{N:{\rm even}}y^N Z_N$,
 where the fugacity
 $y=e^{-\Delta/\kBT}$.
 $\Delta$ is a monopole creation energy, which is, to some extent,
 controlled by ${\bm H}_{\rm dc}$.
 Then,
 we can estimate the asymptotic behavior of the charge correlation
 function as
 $C(r)\simeq-\langle \rho({\bm x}) \rho({\bf 0})\rangle_{\rm CG}$,
 where
 $\langle\cdots\rangle_{\rm CG}$ means an average with respect to
 $\Xi_{\rm CG}$.

 While CG possesses a low-temperature phase ($\kappa\ge4$) where
 monopoles with opposite charges are bounded into pairs
 \cite{Kost74},  
 kagom\'e ice corresponds to CG in the high-temperature phase
 ($\kappa\simeq\frac12$), and thus, exhibits a screened charge
 correlation due to free monopoles.
 To evaluate $C(r)$, we utilize the well-known equivalence between CG
 and the sine-Gordon model
 \cite{Minn87,Poly77}
 defined by the partition function
 $Z_{\rm sG}=\int[d\vartheta]e^{-{\cal S}_{\rm sG}}$
 with the action
 \cite{Otsu11}:
 \begin{equation}
  {\cal S}_{\rm sG}
   =
   \int d^2x
   \Bigl[
   \frac{1}{2\pi\kappa}\left(\nabla\vartheta\right)^2 
   -2z\cos\sqrt{2}\vartheta
   \Bigr],
   \label{sG}
 \end{equation}
 where $z\simeq (y/\zeta)a_0^{\kappa/2}$
 ($\zeta$ is an area of unit cell of $\Lambda$).
 Then, in the sine-Gordon language, $C(r)=4z^2c(r)$ $(r\ne0)$ with
 \begin{equation}
  c(r)
   =
   \langle
    \sin\sqrt{2}\vartheta({\bm x})
    \sin\sqrt{2}\vartheta({\bf 0})
   \rangle_{\rm sG},
   \label{c_sG}
 \end{equation}
 where $\langle\cdots\rangle_{\rm sG}$ means an average with respect to
 $Z_{\rm sG}$.
 In this formulation, $c(r)$ is short-ranged due to the relevant
 nonlinear term in Eq.\ (\ref{sG}).
 In such cases, it can be calculated reliably by using the formfactor
 perturbation (FFP) method. 
 While the method tells us elementary processes to be considered and
 provides a simple expression for $c(r)$, its explanation is devoted to
 a rather technical aspect of the massive sine-Gordon theory.
 Thus, here we only provide the result---for readers interested in
 details, please see the Supplementary Material (SM):
 \begin{equation}
  c(r)\simeq\left(\frac{\lambda\nm}{2z}\right)^2\frac{K_0(m_1r)}{\pi},
   \label{cr}
 \end{equation}
 where $K_\alpha$ denotes the $\alpha$th-order modified Bessel function
 of the second kind.
 Thus, $C(r)$ is short-ranged with $\xi=1/m_1$, where $m_1$ is the mass
 of the principal breather $B_1$.

 Now, we are in a position to obtain the ac susceptibility.
 Performing the integral transform of Eq.\ (\ref{ReX})
 \cite{CALTECH},
 we obtain the real part as
 \begin{equation}
  {\chi'(\omega)}/{\chi(0)}
   \simeq
   1-\frac{\gamma^3}{8}\left[K_1(\gamma)+K_3(\gamma)\right].
   \label{ReX_result}
 \end{equation}
 We defined a characteristic frequency of the $B_1$ excitation (see SM)
 as
 $\omega_1={\cal D}/\xi^2$
 and a ratio as
 $\gamma=bL_\omega/\xi=b\sqrt{\omega_1/\omega}$.
 The static susceptibility
 $\chi$ $[=\chi(0)]$
 which gives a magnitude of $\chi(\omega)$ is found to obey Curie's law:
 \begin{equation}
  \chi(0)
   \simeq
   \Omega
   \frac{4\mu^2_{\rm eff}}{\kBT a^2}\lambda^2\nm^2\xi^4
   =
   \frac{\cal C}{\kBT}.
   \label{X0}
 \end{equation}
 Intriguingly, Curie's constant depends on $\kappa$ characterizing the
 ground-state manifold
 (${\cal C}\simeq\Cu\times\mu^2_{\rm eff}N_{\Lambda^*}$ for
 $\kappa=\frac12$ and $N_{\Lambda^*}$ spins),
 which may give one aspect of the 2D cooperative paramagnets.
 The imaginary part is obtained by the KK relation; the result is
 written, in terms of Meijer's $G$ functions
 \cite{Meij36},
 as
 \begin{equation}
  {\chi''(\omega)}/{\chi(0)}
   \simeq
   \frac{\gamma^2}{8\pi}
   \Bigl[
   \MeijerG{\frac{\gamma^4}{2^8}}{{\bf a}}-
   \MeijerG{\frac{\gamma^4}{2^8}}{{\bf b}}\Bigr],
   \label{ImX}
 \end{equation}
 where vectors
 ${\bf a}=(0,0,1,\frac32,-\frac12)$
 and
 ${\bf b}=(0,0,\frac12,1,\frac12)$.

 Before exploring ingredients of Eqs.\ (\ref{ReX_result})--(\ref{ImX}),
 two comments are in order: 
 (i)
 Because we have focused in the vicinal region from the ground-state
 manifold of kagom\'e ice, the temperature should not be so high to
 bring the system out of the region.
 Moreover,
 ${\bm H}_{\rm dc}$ should be at least in the interval to give the
 magnetization plateau.
 (ii)
 The FFP expansion provides an efficient approximation for the
 long-distance behavior of $c(r)$, but it is not for the short-distance
 one.
 Despite of this fact, we expect it to work at least in the
 low-frequency (i.e., long-distance) region because the $r^3$ factor in
 Eq.\ (\ref{ReX}) relatively amplifies the contribution from the
 long-distance part of the charge correlation.

 \ZUni

 In Fig.\ \ref{acX}, we give Eqs.\ (\ref{ReX_result}) and (\ref{ImX}) by
 the solid lines.
 Since $\omega$ is always scaled by $\omega_1$, these curves are
 independent of model parameters.
 However, 
 it is, strictly speaking, due to the single-mode approximation in the
 FFP expansion; thus, the corrections to scaling may possibly be
 expected.
 In this plot, one finds that $\chi'(\omega)$ exhibits a steep decrease
 at around $\omega_1$, where $\chi''(\omega)$ forms an asymmetric peak.
 Therefore,
 $B_1$ is responsible for the in-phase component, and its delay in
 response against ${\bm H}_{\rm ac}(t)$ causes a dephasing, which is
 then detected as the peak in $\chi''(\omega)$.
 The frequency $\omega_1$ is the square of the $B_1$ mass, so it
 increases as a power of $z$ with the exponent $1+p$ (see SM).
 Thus,
 with increasing $z$, but keeping $T$ constant, the peak shifts toward
 higher frequency region according to the power law, but its height is
 almost unchanged (if ${\cal D}$ is a constant).
 We expect that these behaviors as well as the scaling properties give a
 hallmark of the ac magnetic response observed in the 2D spin ice.

 Finally,
 we perform Monte Carlo (MC) simulations of kagom\'e ice by using the NN
 model of $H_{\rm ice}$ with $J_{\rm eff}=4.4$\ K
 \cite{Harr97,Taba06,Melk01}.
 We employ a system of $54\times54$ pairs of tetrahedra, and simulate it
 at $|{\bm H}_{\rm dc}|=0.3$\ T and $T=0.6$, 0.7 and 0.8\ K, where {\Dy}
 is known to be in the kagom\'e-ice state
 \cite{Taba06}.
 We estimate $\nm$ and $C(r)$ by MC simulations. 
 Since the sublattice dependence of $q_l$ in the lattice model hinders
 detection of large-scale behavior of $C(r)$, we performed a coarse
 graining of charge distributions using the unit cell of $\Lambda$.
 We summarize our simulation results in Fig.\ \ref{acX}(a).
 One can find that scaled data of different $T$ (and thus $\nm$) given
 by marks with error bars agree with the theoretical curve.
 This fact suggests that our theory is applicable for the ac magnetic
 response in kagom\'e ice.
 The inset of Fig.\ \ref{acX}(a) gives the $bL_\omega$ dependence of
 $\chi'$ defined by Eq.\ (\ref{ReX}), which exhibits a saturation to a
 static value $\chi(0)$.
 Further, $\xi$ can be evaluated from $\nm$ (see SM).
 Therefore, we can plot the MC simulation data as functions of
 $\omega/\omega_1$ $[=(\xi/L_\omega)^2]$.
 In the inset of Fig.\ \ref{acX}(b), we plot $1/\chi(0)$ versus $T$.
 Although we have predicted Curie's law, a small deviation is visible.
 This may be due to the finite-size effects in low $T$ region, but more
 intensive numerical simulations are necessary for the kagom\'e-ice
 model.

 In conclusion, we have investigated the ac susceptibility
 $\chi(\omega)$ of kagom\'e ice: 
 We clarified that $\chi(\omega)$ takes an universal scaling form in
 terms of the ratio of $\omega$ to $\omega_1$, the characteristic
 frequency for the principal breather---
 it is a localized excitation composed of a soliton and antisoliton, but
 possibly looks more similar to an ordinary wave
 \cite{Sasa86}.
 Furthermore, we performed MC simulations, and provided data that
 represent the scaling form as expected in our theory.
 The present results suggest that breather's dynamics characterizes the
 low-$T$ behavior of the magnetic monopoles in kagom\'e ice.
 In this letter,
 it
 has been explained that the universal dynamics of the magnetic
 monopole-like defects in the 2D spin ice can be captured by the theory
 of the ac magnetic response.
 Now, in view of the universality concept, it is natural to expect that
 our theory can also serve for analysis of the dynamics observed in
 other systems such as the vortices in superfluid and superconducting
 films, as well as in 2D XY magnets above the Kosterlitz-Thouless
 transition, and also the charged particles in 2D electrolytes.
 These remain as interesting future applications.

 The authors thank
 Y. Okabe, 
 G. Tatara,
 M. Fujimoto, 
 A. Tanaka, 
 and 
 K. Nomura
 for stimulating discussions.
 Main computations were performed using the facilities of Cyberscience
 Center in Tohoku University.

 \newcommand{\AxS}[1]{#1,}
 \newcommand{\AxD}[2]{#1 and #2,}
 \newcommand{\AxM}[2]{#1, and #2,}
 \newcommand{\REF }[4]{#1 {\bf #2}, #3 (#4)}
 \newcommand{\JPSJ}[3]{\REF{J. Phys. Soc. Jpn.\           }{#1}{#2}{#3}}
 \newcommand{\PRL }[3]{\REF{Phys. Rev. Lett.\             }{#1}{#2}{#3}}
 \newcommand{\PRA }[3]{\REF{Phys. Rev.\                  A}{#1}{#2}{#3}}
 \newcommand{\PRB }[3]{\REF{Phys. Rev.\                  B}{#1}{#2}{#3}}
 \newcommand{\PRE }[3]{\REF{Phys. Rev.\                  E}{#1}{#2}{#3}}
 \newcommand{\PRX }[3]{\REF{Phys. Rev.\                  X}{#1}{#2}{#3}}
 \newcommand{\NPB }[3]{\REF{Nucl. Phys.\                 B}{#1}{#2}{#3}}
 \newcommand{\JPA }[3]{\REF{J. Phys.\ A: Math. Gen.       }{#1}{#2}{#3}}
 \newcommand{\JPC }[3]{\REF{J. Phys.\ C: Solid State Phys.}{#1}{#2}{#3}}
 \newcommand{\IBID}[3]{\REF{{\it ibid.}}{#1}{#2}{#3}}

\clearpage
\newcommand\Ma{6.22}
\newcommand\La{1.21}
\setcounter{equation}{0}
\def\theequation{S\arabic{equation}}
\widetext
\begin{center}
 {\large\bf Supplementary Material:\\ ``\TITLE''}
\end{center}

 This Supplementary Material contains an explanation on the 
 formfactor perturbation (FFP) calculation of the charge correlation
 function defined by Eq.\ (\ref{c_sG}) and its lowest-order result given
 by Eq.\ (\ref{cr})
 \cite{Smir92}. 
 Also, a relationship between the charge correlation length and the
 defect number density is obtained as a by-product.

 The model Eq.\ (\ref{sG}) possesses low-energy excitations of the soliton
 $s$, the antisoliton $\bar s$, and breathers $B_j$ for $0\le\kappa<2$
 ($j=1,\cdots,\lfloor 1/p\rfloor$ with $1/p=4/\kappa-1$)
 \cite{Fadd74,Zamo79}.
 The mass spectrum $m_\epsilon$ consists of a doublet of $s$ and
 $\bar s$, $m_\pm=M$, and singlets of $B_j$,
 \begin{equation}
  m_j=2M\sin\left(\frac{\pi p}{2}j\right).
 \end{equation}
 The soliton mass varies as a power of the scaling field $z$
 \cite{Dest91,Zamo95}:
 \begin{equation}
  M
   =
   \frac{2\Gamma(p/2)}{\sqrt\pi\Gamma((1+p)/2)}
   \left[\pi\frac{\Gamma(1/(1+p))}{\Gamma(p/(1+p))}z
   \right]^{\frac{1+p}{2}}, 
 \end{equation}
 and represents an inverse length scale
 ($M\simeq\Ma\times z^{\frac47}$ for $\kappa=\frac12$).
 The FFP method expands the correlation function as
 $c(r)=\sum_{{\cal N}=1}^\infty c_{\cal N}(r)/{\cal N}!$,
 where the contribution from the ${\cal N}$-excitation sector is given
 by 
 \begin{equation}
  c_{\cal N}(r)
   =
   \prod_{k=1}^{\cal N}
   \Bigl[
   \sum_{\epsilon_k}
   \int_{-\infty}^\infty
   \frac{d\theta_k}{2\pi}
   e^{-E_k(\theta_k)r}
   \Bigr]
   \Bigl|F_{\sin}(\{\theta\})_{\{\epsilon\}}\Bigl|^2.
 \end{equation}
 The two sets
 $\{\epsilon\}=\{\epsilon_1,\cdots,\epsilon_{\cal N}\}$
 and 
 $\{\theta\}=\{\theta_1,\cdots,\theta_{\cal N}\}$
 specify aforementioned species of excitations and their rapidities,
 respectively. 
 An $\epsilon_k$ excitation with a rapidity $\theta_k$ has the energy
 $E_k(\theta_k)=m_{\epsilon_k}\cosh\theta_k$.
 The formfactor,
 $F_{\sin}(\{\theta\})_{\{\epsilon\}}$,
 represents a matrix element of
 $\sin\sqrt{2}\vartheta$
 between the ground state and excited states, and selects relevant
 excitations to the correlation function.
 In this respect, the invariance of
 Eq.\ (\ref{sG}),
 under the charge conjugation
 ${\cal C}:$\ $\vartheta\to-\vartheta$
 has the central importance
 \cite{Essl97}.
 Since the charge-density operator transforms as
 ${\cal C}:$\ $\sin\sqrt{2}\vartheta\to-\sin\sqrt{2}\vartheta$,
 nonvanishing contributions stem from excitations with odd parity.
 Consequently,
 the leading contribution comes from the principal breather $B_1$
 which is taken into account in the expansion, i.e.,
 ${\cal N}=1$ and $\epsilon_1=1$ in Eq.\ (S3).
 The formfactor is then independent of the rapidity, and is given by
 $F_{\sin}(\theta_1)_1
 =
 -\lambda\langle e^{i\sqrt{2}\vartheta}\rangle_{\rm sG}
 =
 -\lambda\nm/2z$
 \cite{Luky95},
 where the constant
 \begin{equation}
  \lambda
   =
   2\cos\frac{\pi p}{2}
   \sqrt{2\sin\frac{\pi p}{2}}
   \exp\Bigl(-\int_0^{\pi p} \frac{tdt}{2\pi\sin t}\Bigr),
 \end{equation}
 and the defect number density
 \cite{Luky97,Sama00}
 \begin{equation}
  \nm=\frac{(1+p)}{4}\tan\left(\frac{\pi p}{2}\right)M^2.
 \end{equation}
 As a result,
 the charge correlation function is simply given by
 Eq.\ (\ref{cr}).
 Since its asymptotic behavior in $r\to\infty$ can be written as
 $c(r)\propto \exp(-m_1r)/\sqrt{r}$, the charge correlation length
 $\xi=1/m_1$.
 Therefore, from Eqs.\ (S1) and (S5), we obtain the 
 relationship between $\xi$ and $\nm$ as
 \begin{equation}
  \xi^{-1}=\sqrt{\frac{8\nm}{1+p}\sin(\pi p)}. 
 \end{equation}
 For instance, 
 $\xi^{-1}=\sqrt{7\nm\sin(\pi/7)}$ for $\kappa=\frac12$.

\end{document}